\documentclass[amssymb,prb,twocolumn,superscriptaddress,floats,showkeys,showpacs]{revtex4-1}
\usepackage[T1]{fontenc}
\usepackage{bm}
\usepackage{graphicx}
\usepackage{amssymb}
\usepackage{leftidx}
\usepackage{upgreek}
\usepackage{siunitx}
\usepackage{amsfonts}
\usepackage{amsmath} 
\usepackage{hyperref} 
\usepackage{comment}
\usepackage{afterpage}
\hypersetup{colorlinks,citecolor=blue, filecolor=blue ,linkcolor=blue , urlcolor=blue, pdftex}

\usepackage{tabularx}

\begin{document}

\title{Ion-beam-milled graphite nanoribbons as mesoscopic carbon-based polarizers.}


\def \FUW{Institute of Experimental Physics, Faculty of Physics, University
of Warsaw, Pasteura 5, 02-093 Warsaw, Poland}

\author{Marcin Muszyński} \affiliation{\FUW}
\author{Igor Antoniazzi} \affiliation{\FUW}
\author{Bruno Camargo} \email[]{b.c_camargo@yahoo.com.br} \affiliation{\FUW}

\begin{abstract} 
We demonstrate optical reflectivity and Raman responses of graphite microstuctures as a function of light polarization when the incident light is applied perpendicular to the material's stacking direction (c-axis). For this, we employed novel graphite nanoribbons with edges polished through ion-beam etching. In this unique configuration, a strong polarization dependence of the D, G and 2D Raman modes is observed. At the same time, polarized reflectivity measurements demonstrate the potential of such a device as a carbon-based, on-chip polarizer. We discuss advantages of the proposed fabrication method as opposed to the mechanical polishing of bulk crystals.
\end{abstract}

\maketitle


Graphite is a highly anisotropic carbon-based material composed of a stack of weakly-bonded graphene layers in a Bernal (ABAB) structure \cite{Chung2002}. Usually encountered in nature as a highly-oriented crystal, this material has been on the forefront of technological advances since its discovery. Recently, it has found applications in electronics as few- and multi-layer graphene on tentative detectors \cite{Bonaccini2022, Wasfi2022} and battery electrodes \cite{Asenbauer2020}.

The crystallographic structure of graphite makes it naturally cleavable along the stacking direction, dubbed as the c-axis. However, because single crystals have yet to be achieved, cutting in any other direction usually poses a more arduous task. For this reason, physical property measurements usually focus on in-plane electrical, magnetic and optical properties, with off-plane measurements often being strongly sample dependent (see, e.g. ref. \cite{Camargo2016} and the SI).
Among other reasons, this is caused by stacking faults in the highly oriented structure (which both hinders c-axis transport and couples the dispersion of $k_z$ to the plane \cite{Matsui2022}), and defects along the edges of the sample. The latter are uncleavable and present irreproducible irregularities when cut.

\begin{figure}[b]
    \centering
    \includegraphics[width=1\linewidth]{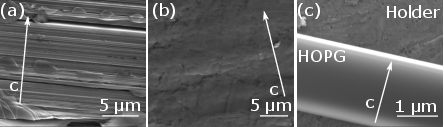}
    \caption
    {\label{fig:1} SEM image of the edges of a diamond-saw-cut bulk HOPG crystal in a) a region where the sample was broken, b) in a mechanically-polished region and c) in a ion-beam-polished ribbon. The arrows point the c-axis direction.}
\end{figure}

These irregularities make the measurement of optical properties of graphite cross-sections particularly challenging. Mechanical polishing of the edge of the samples is bound to cause graphite and graphene flakes to bend (see Fig. \ref{fig:1}) and cleave out \cite{Jellison2007}, introducing an in-plane response to an otherwise out-of-plane measurement. This results in a limited number of literature reports on the subject. 

Among the properties seldom addressed are in-plane light polarizability and Raman spectra. 
Although careful measurements performed in the 1980s and 1990s addressed such topics to a certain extent, polishing techniques used at the time resulted in large in-plane contributions to the measured data, tarnishing the results \cite{Katagiri1988, Kawashima1999, Jellison2007}. More recently, attempts to bypass this issue have been made by Tan et al. \cite{Tan2004} by employing a pristine, mined natural graphite crystal. Although such an approach intended to mitigate the effects of polishing the surface, the soft properties of graphite do not allow for an edge without bent over planes.

Here, we revisit this problem, focusing on Raman spectroscopy of graphite when the light incides parallel to the graphene planes. However, instead of using mechanically polished bulk Highly oriented pyrolitic graphite (HOPG) as in refs \cite{Katagiri1988, Kawashima1999}, we employ nanoribbons fabricated through ion-beam milling. This type of samples have been previously used to study the impact of stacking faults on the electrical transport properties of graphite \cite{Ballestar2013}. However, such devices could also be used to probe other off-plane phenomena in HOPG, e.g. ionic channeling \cite{Camargo2017} and optical properties. This occurs because graphite samples fabricated this way are ribbons whose sides are covered by a thin (approx. 30 nm), yet transparent layer of amorphous carbon, while minimizing the contributions arising from graphitic planes bent from the edges \cite{Ballestar2013} (which occur in mechanical polishing methods).  Unlike previous attempts to measure off-plane optical properties of graphite, our approach also has the potential to permit the fabrication of optically transparent samples, in principle enabling one to probe the properties of both reflected and transmitted light through graphite ribbons.

The latter is important because, due to the ever-increasing demand of higher processing speeds at lower power consumption, the employment of photonic integrated circuits has become a necessity for the advancement of the semiconducting industry. Among the studied systems, graphene and the possibility of its utilization in valley- and spin-tronics has contributed to the increased interest in the interplay between light and electrical transport in mesoscale devices \cite{Mrudul2021, Higuchi2017, Rong2023}, in a bid to utilize chiral or valley-selected electrons as quantum-information-carrying agents. Proposals introduced in the last decade to miniaturize and integrate conventional electronics with optics involve the implementation of modulators, couplers and filters in reduced scale, both based on graphene \cite{Mrudul2021}, conventional semiconductors \cite{Liu2021}, or hybrids of these two materials \cite{Xie2018}. However, such structures are sensitive to light polarization due to the structural birefringence of the waveguides utilized. Therefore, to properly employ such circuits, it is necessary to also introduce mesoscopic polarizers to filter out unwanted radiation components. Among the most common types of reflection polarizers utilized for such a task, are Bragg gratings and photonic crystals, which are complex structures \cite{Goldstein2003}. We present an alternative, using a device that can be etched from a monolithic block of graphite, thus being completely compatible with graphene-based circuits.

The samples used in this work consisted of thin graphite ribbons with approximate dimensions of 10 $\mu$m $\times$ 300 nm (ab-direction) $\times$ 5 $\mu$m (c-axis). They were prepared with ion-milling etching using a FEI dual beam microscope, following the procedure previously outlined in refs. \cite{Camargo2017, Barzola2011}. Two devices were fabricated with different milling currents. This warranted specimens coated with a thin layer of amorphous carbon, with thicknesses estimated at 30 nm (Sample A) and 60 nm (Sample B) (estimated with SRIM \cite{SRIM}). The graphite ribbons were then deposited atop a mirror-polished $\text{SiN}_2$ substrate, with graphite's c-axis aligned parallel to the $\text{SiN}_2$ surface (graphite planes were perpendicular to the substrate's surface). Afterwards, samples were optically characterized by means of polarization-sensitive Raman spectroscopy and by reflectivity measurements in the visible range. The incident light was always applied perpendicular to the graphite c-axis.

Polarization-sensitive Raman scattering (RS) experiments were performed in the backscattering geometry with 515~nm excitation, using a long-working-distance 50× objective  with numerical aperture 0.55. The beam spot size was of approx. 1~$\mu$m. and was aimed at the center of the sample. The scattered light was sent through a 0.75~m monochromator, and collected by a charge-coupled device (CCD) cooled by liquid nitrogen.

Reflectivity measurements were performed by illuminating the sample with a non-polarized white light source. A 50x microscope objective was used to focus and collect light from a spot with a diameter of approx. 50~$\mu$m. The spot was centered at the HOPG sample, but also probed a clean region on the substrate, which was later used as a reference. A linear polarizer was placed on a motorized rotation stage to detect the polarization of the collected light. The sample image was focused by a f~=~100~mm lens on a CCD~camera. All measurements were performed at room temperature.

\begin{figure}[ht]
    \centering
    \includegraphics[width=1\linewidth]{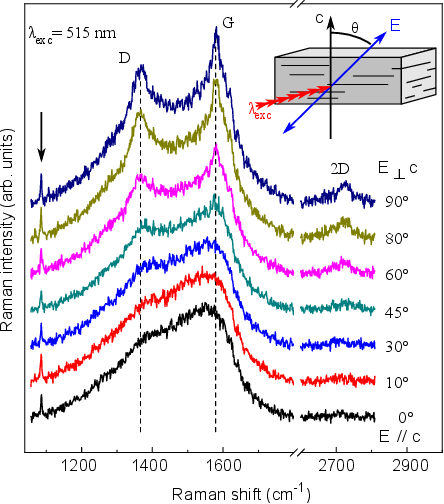}
    \caption
    {\label{fig:2} Colinear Raman spectra obtained for different polarizations of the excitation light. The curves have been displaced vertically for clarity. The inset shows a cartoon of the measurement geometry; $\theta = 0^\circ$ corresponds to the excitation polarized along the sample's c-axis. The main graphite modes, labelled G, D, and 2D, are extinguished as the excitation is polarized along the sample's c-axis. The broad background and the peak at 1087 $\text{cm}^{-1}$ (indicated by an arrow) are weakly-dependent on $\theta$.}
\end{figure}

Raman spectra of the graphite ribbon labelled as Sample A are presented in Fig.~\ref{fig:2} for different values of $\theta$, defined as the angle between the polarization of the incident light and the sample's c-axis. Results for E$\bot$c ($\theta= 90^\circ$, electric field aligned in-plane) revealed the presence of bands at $1366 $ $\text{cm}^{-1}$ and $1580 $ $\text{cm}^{-1}$ , whose intensity diminished upon changing the light polarization towards the out-of-plane direction E$\parallel$c ($\theta = 0^\circ$). This result confirms these modes as due to the in-plane displacement of carbon atoms \cite{Jellison2007}. Indeed, such lines have been reported by many authors \cite{Ferrari2007, Katagiri1988, Kawashima1999, Jellison2007, Tan2004}, and correspond to the in-plane D and G modes of graphite, respectively. The former has been attributed to a defect-induced inter valley double resonance process, whereas the latter is associated with a ``stretching'' mode centered around k = 0 \cite{Ferrari2007}. 

Superimposed to such a spectra was a broad background, which did not show any dependence on the incident polarization, thus suggesting its origin as a result of an amorphous layer covering the sample. The positions of their maxima are characteristic of a carbonic material with an intermix of $\text{sp}^2$ and $\text{sp}^3$ bonding \cite{Pimenta2007}.

 Curiously, we report a previously unobserved feature at 1087 $\text{cm}^{-1}$. This sharp peak did show an isotropic dependence during the measurements. Upon closer inspection, its presence was also found in database Raman spectra of mineral natural graphite from different sources~\cite{RRUFF} - hailed by many as a material of superior quality~\cite{Precker2019, Zhu2019, Edmann1998, Pendrys1980}. However, to date, no similar observations have been reported in in- and out-of-plane measurements of synthetic HOPG~\cite{Katagiri1988, Kawashima1999, Tuinstra1970, Tan2004}. We tentatively attribute this mode to the stretching vibrational modes of highly ordered H-terminated =C-C=C- chains along the edges of graphite, akin to 1D trans-polyacetylene \cite{Paraschuk2002}. In such a compound, the peak at 1087~$\text{cm}^{-1}$ is usually accompanied by a feature at 1470~$\text{cm}^{-1}$, which also coincides with the broad $\theta$-independent maximum superimposed to our G and D peaks. Its existence in our system can be justified in the context of hydrogen-passivated zigzag edges of graphite, which is usually grown from hydrogen-rich precursors and always contains a large content of H$_2$ molecules near the surface~\cite{Ohldag2010, Reichart2006}. We hypothesize that the ion milling-induced decomposition of the sample surface also decomposes interlayer adsorbed H$_2$, similar as reported for proton-irradiated graphite~\cite{Ohldag2010}.  This results in passivized graphite edges below an amorphous carbon layer, as opposed to a plethora of different carbon-carbon orbital terminations expected in mechanically-polished graphite. 

We stress that the latter method is employed by the entirety of reports in the literature in order to gain access to graphite's out-of-plane optical properties (see e.g. refs~\cite{Katagiri1988, Kawashima1999, Tuinstra1970, Tan2004}). Yet, the low stiffness and layered structure of the material make such a mechanical process challenging, invariably resulting in edge structures similar to those shown in Fig. \ref{fig:1}, containing irregular and/or bent-over flakes that contribute to the measured signal of macroscopic specimes. Thus, the fundamental difference between our samples and those found in the literature: the edges of the samples shown here were polished in a contact-free ion-milling process, erasing the possibility of foldings and exfoliations at their edges, caused by friction.

\begin{figure}[h]
    \centering
    \includegraphics[width=1\linewidth]{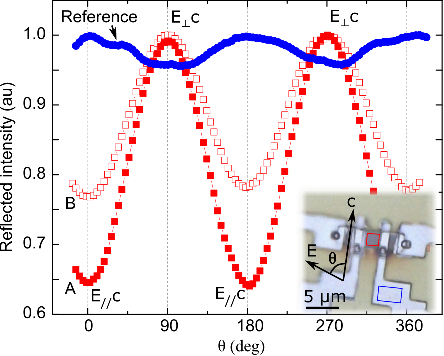}
    \caption
    {\label{fig:3} Normalized reflected light intensity as a function of detection polarization for a region in the graphite sample (square symbols) and a region on the metal deposited atop the $\text{SiN}_2$ substrate (round symbols). Closed squares correspond to a graphite ribbon covered by a ca. 30 nm amorphous C layer (sample A), whereas empty squares correspond to the data of a sample covered by a ca. 60 nm amorphous C layer (sample B). The inset shows a picture of sample A, with the regions where the light intensity was probed indicated by squares.}
\end{figure}

We demonstrate the behavior of our device as a small-scale on-chip polarizer. Reflectance measurements, presented in Fig.~\ref{fig:3}, show the intensity of the reflected light in our device as a function of the linear polarization angle with respect to the sample's c-axis. The intensity of the reflected light in the configuration E$\bot$c is approx. 65\% larger than for $E||c$. The reflectivity spectra measured in visible range for sample A for two orthogonal polarization are presented as a degree of linear polarization in the SI (Fig.~S5). The ratio between the reflected intensities in both polarizations, however, strongly depended on the thickness of the amorphous layer covering the sample. Data obtained for device B - in which a 60~nm amorphous layer was expected - presented a polarization ratio of approx. 42\%. These results suggest that the amorphous carbon layer obtained during milling acts as a diffusor, whose efficiency can be selected by controlling its thickness during ribbon preparation. 

Indeed, Raman measurements performed on device B did not present as sharp G and D lines as in device A for E$\bot$c (see the SI). Such a result attests to a higher surface amorphization, as well as an enhanced scattering of light by the amorphous carbon layer coating the sample.

 \vfill\null

The polarization observed in reflectivity has its origins in the high anisotropy of the graphite conductivity tensor. Indeed, elipsometry measurements dedicated to probe the ordinary and extraordinary dielectric constants of graphite did reveal a non-zero imaginary component of the extraordinary $\varepsilon$, characterizing the material as a conventional metal for electric fields parallel to the planes (E$\bot$c), and as a semiconductor or semimetal otherwise \cite{Jellison2007}. We note, however, that experiments dedicated probe such properties are few and far between, and might yield conflicting results \cite{Jellison2007, Papoular2014}. This occurs mainly because of the difficulty in controlling the roughness of surfaces parallel to graphite's c-axis through mechanical polishing. Our results suggest that such an issue might be partially addressed by employing the ion-beam milled samples showcased here.

Although presenting a small degree of polarization (ca. 0.25 for the best sample), it is conceivable that the quality of the polarizing devices presented in Fig. \ref{fig:3} can be improved upon by thinning the amorphous layer at the surface of graphite. Different approaches are currently available to tackle this issue, such as slow cleaning of the sample with low-intensity oxygen or argon plasma, as utilized e.g. in ref. \cite{Precker2019} for the electrical addressing of graphite along its stacking direction. The tracking of the optical properties of graphite as a function of the thickness of the amorphous coating will be discussed elsewhere.

Because the fundamental optical properties of pristine graphite ribbons explored here rely on its high electrical anisotropy, we envision our devices as components of miniaturized all-optical contact-less detectors responsive to any effects disturbing the stacking order on graphite. Among them are the sensing of traditional graphite interstitials such as Li and H \cite{peng2020, An2019}. A graphite-based optical device milled from a monolithic block of graphite would, thus, have the potential to act as a fully-contained optical chemical sensor to probe, for example, charge in Li-based batteries or H diffusion in hydrogen cells. Because our devices are conducting, they can also act as electrodes for electric-field-tunable carbon-based polarizers, such as azo-group liquid crystal \cite{Meng2016, Zhang2020}, effectively enabling an all-carbon method to locally control the circular polarization of light. Such a technology is fundamental for the implementation of valleytronics in integrated devices, whose carrier population selection currently relies on optical pumping with circularly polarized light \cite{Rong2023, Li2020, Navdeep2023}.




In summary, we demonstrate the anisotropic Raman and reflectivity properties of thin graphite ribbons prepared through ionic etching, in the unusual configuration when the incident light is perpendicular to the sample c-axis. Results revealed that our devices act as small-scale polarizers for the reflected light. This is caused by the better coupling between electric fields and electronic motion when light is polarized along the planes (E$\bot$c), as opposed to parallel to the c-axis direction (E$\parallel$c). This result, while seemingly unsurprising, demonstrates the applicability of small graphite structures for on-chip optical all-carbon-based devices, resulting in small components chemically identical to graphene. The samples considered here had a thickness of approx. 300 nm. However, state-of-the-art ion-beam-milled HOPG lamellae prepared for transmission electron microscopy measurements (e.g., as employed in ref. \cite{Esquinazi2014}) are currently achievable with thicknesses below 100 nm, while still maintaining graphite's stacking structure intact. Recent advancements in single-crystal graphite production, as reported in ref. \cite{Zhang2022}, also hint at the possibility of achieving similar devices through cleaving along high symmetry directions other than graphite's c-axis. In these devices, we expect measurements of transmitted light to be possible, with a similar outcome as the one reported here for the reflected light. We leave such an investigation for a future work. 

\section*{Supplementary Material}
The supplementary information displays additional spectroscopic data, results for sample B, and compares samples of different origins.

\section*{acknowledgements}
We thank Nilesh Dalla (FUW) for assistance with SEM imaging and Jacek Szczytko (FUW) for helpful discussions. We gracefully acknowledge Adam Babiński and Maciej Molas for providing access to the facilities of the Laboratory of Optical Spectroscopy (LaSsO) at the Faculty of Physics, University of Warsaw. This work has been supported by the National Science Centre, Poland (grants no. 2017/27/B/ST3/00205, 2019/35/B/ST3/04147 and 2018/31/B/ST3/02111).

%
4

\end{document}